\documentclass[fleqn,10pt]{wlscirep}

\title{Memristive Sisyphus circuit for clock signal generation}
\author[1,2,3,*]{Yuriy V.~Pershin}
\author[4,5,2]{Sergey N.~Shevchenko}
\author[2,6]{Franco~Nori}

\affil[1]{Department of Physics and Astronomy and Smart State Center for Experimental Nanoscale Physics, University of South Carolina, Columbia, South Carolina 29208, USA}
\affil[2]{CEMS, RIKEN, Saitama 351-0198, Japan}
\affil[3]{Nikolaev Institute of Inorganic Chemistry SB RAS, Novosibirsk 630090, Russia}
\affil[4]{B.~Verkin Institute for Low Temperature Physics and Engineering, Kharkov 61103, Ukraine}
\affil[5]{V.~Karazin Kharkov National University, Kharkov 61022, Ukraine}
\affil[6]{Physics Department, University of Michigan, Ann Arbor, Michigan 48109-1040, USA}
\affil[*]{pershin@physics.sc.edu}

\begin{abstract}
Frequency generators are widely used in electronics. Here, we report the design
and experimental realization of a memristive frequency generator
employing a unique combination of only digital logic gates, a single-supply voltage
and a realistic threshold-type memristive device. In our circuit, the oscillator frequency and duty cycle
are defined by the switching characteristics of the memristive device and
external resistors. We demonstrate the circuit operation both experimentally,
using a memristor emulator, and theoretically, using a model memristive
device with threshold. Importantly, nanoscale realizations of memristive devices offer
small-size alternatives to conventional quartz-based oscillators. In
addition, the suggested approach can be used for mimicking some cyclic
(Sisyphus) processes in nature, such as \textquotedblleft dripping ants\textquotedblright or drops from leaky
faucets.
\end{abstract}

\begin{document}

\flushbottom
\maketitle


\section*{Introduction}

\begin{figure*}[th]
\centering
\includegraphics[width=.40\columnwidth]{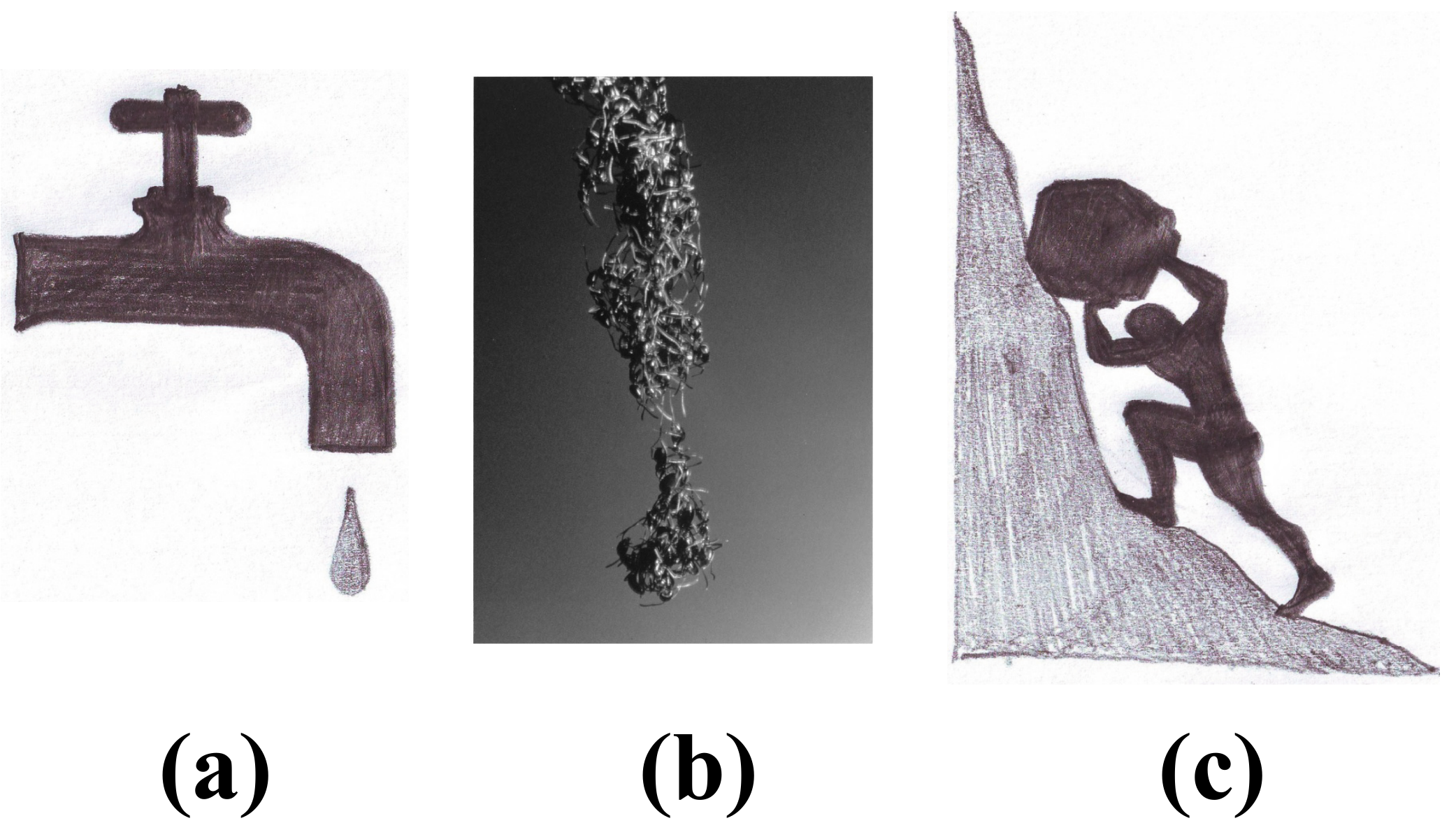} \quad %
\includegraphics[width=.45\columnwidth]{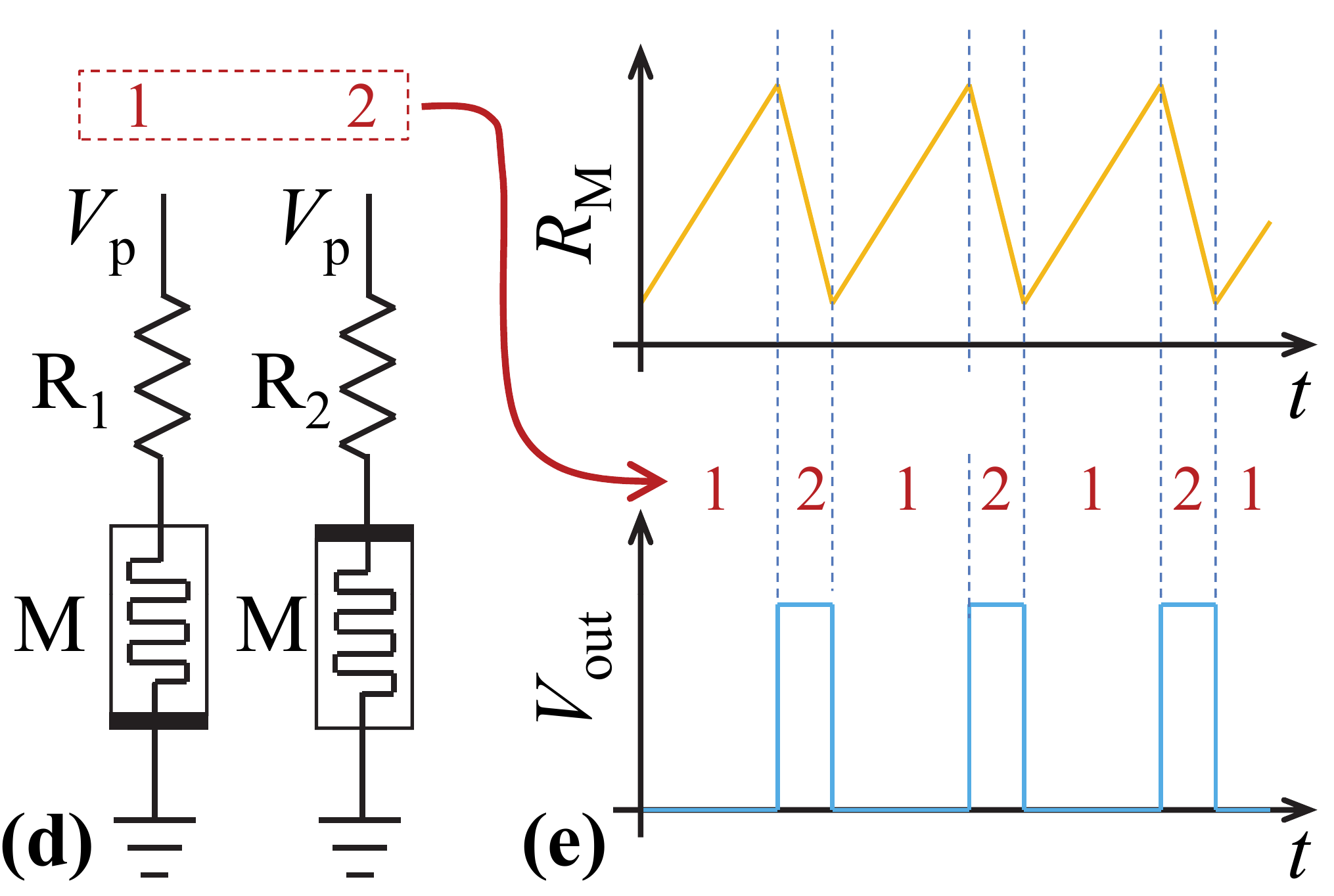}
\caption{(Color online) (a)-(c) Examples of Sisyphus cycles: (a) leaky
faucet, (b) dripping ants \protect\cite{Bonabeau98a,theraulaz2001model}, (c)
mythological Sisyphus. (d) Simplified effective circuits realizing a
two-phase memristive Sisyphus circuit: the increasing memristance stage 1
(left circuit) and decreasing memristance stage 2 (right circuit).
Memristance oscillations and clock pulses are shown schematically on the
graphs (e). (b) is reprinted with permission from Ref.~
\citeonline{Bonabeau98a}.}
\label{fig1}
\end{figure*}

Cyclic evolutions and relaxation oscillators are ubiquitous in nature. One
example of these is the leaky faucet [Fig.~\ref{fig1}(a)], where the
suspended fluid mass increases gradually, until it suddenly decreases, when
the droplet tears off \cite{Utada07,Ambravaneswaran00,Plourde93a}. Other
examples of relaxational dynamics can also be found in granular media \cite%
{Benza93a} as well as in mechanical \cite{Field95b} and superconducting
systems \cite{Field95a}. Similar dynamics can be even found in wildlife. For
example, some types of ants \cite{Bonabeau98a,theraulaz2001model} can
continuously climb a rod, aggregate, and eventually drop, right after a
critical mass is accumulated [see Fig.~\ref{fig1}(b)].

Analogous phenomena in driven dissipative systems are known as Sisyphus
processes. According to Greek mythology, King Sisyphus was doomed to
repeatedly push a rock uphill, which would then roll back down [Fig.~\ref%
{fig1}(c)]. Recently, Sisyphus processes \cite{cohen1990new} were studied in
electric circuits based on superconducting \cite%
{grajcar2008sisyphus,nori2008superconducting,Skinner10a,Persson10a} and
normal-state \cite{gonzalez2015probing,shevchenko2015delayed,Gullans15a}
systems. In these circuits, a driven artificial atom, a qubit, was coupled
to either a mechanical or an electrical resonator. Depending on its
detuning, the qubit was driven by the resonator either \textquotedblleft
uphill\textquotedblright\ (the usual Sisyphus process) or downhill (unusual,
or \textquotedblleft happy Sisyphus\textquotedblright\ process). The cycle
was completed by relaxation to the ground state in the presence of an
additional periodic signal.

\begin{figure*}[th]
\centering
\includegraphics[width=.42\columnwidth]{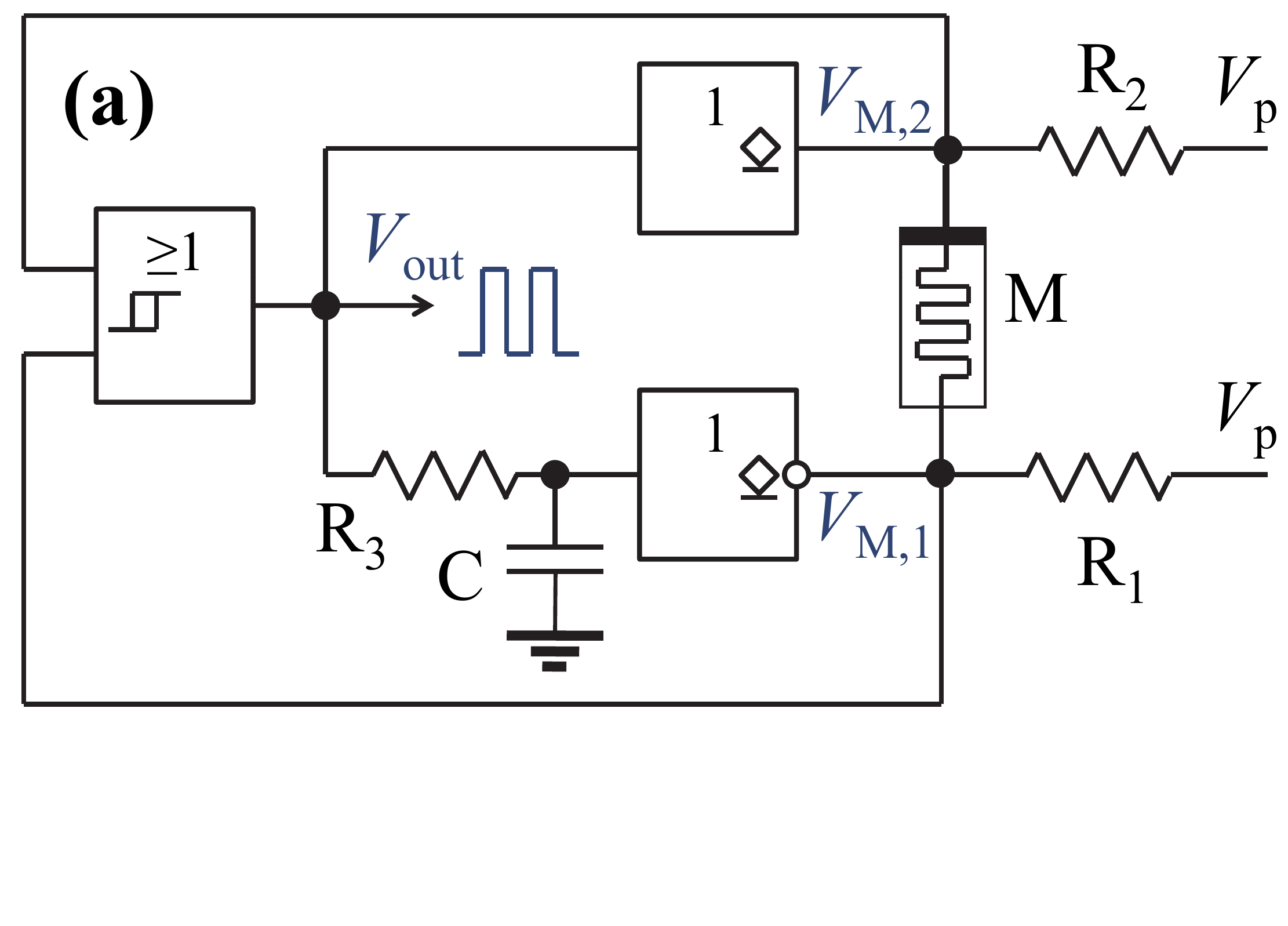} \quad \quad \quad %
\includegraphics[width=.42\columnwidth]{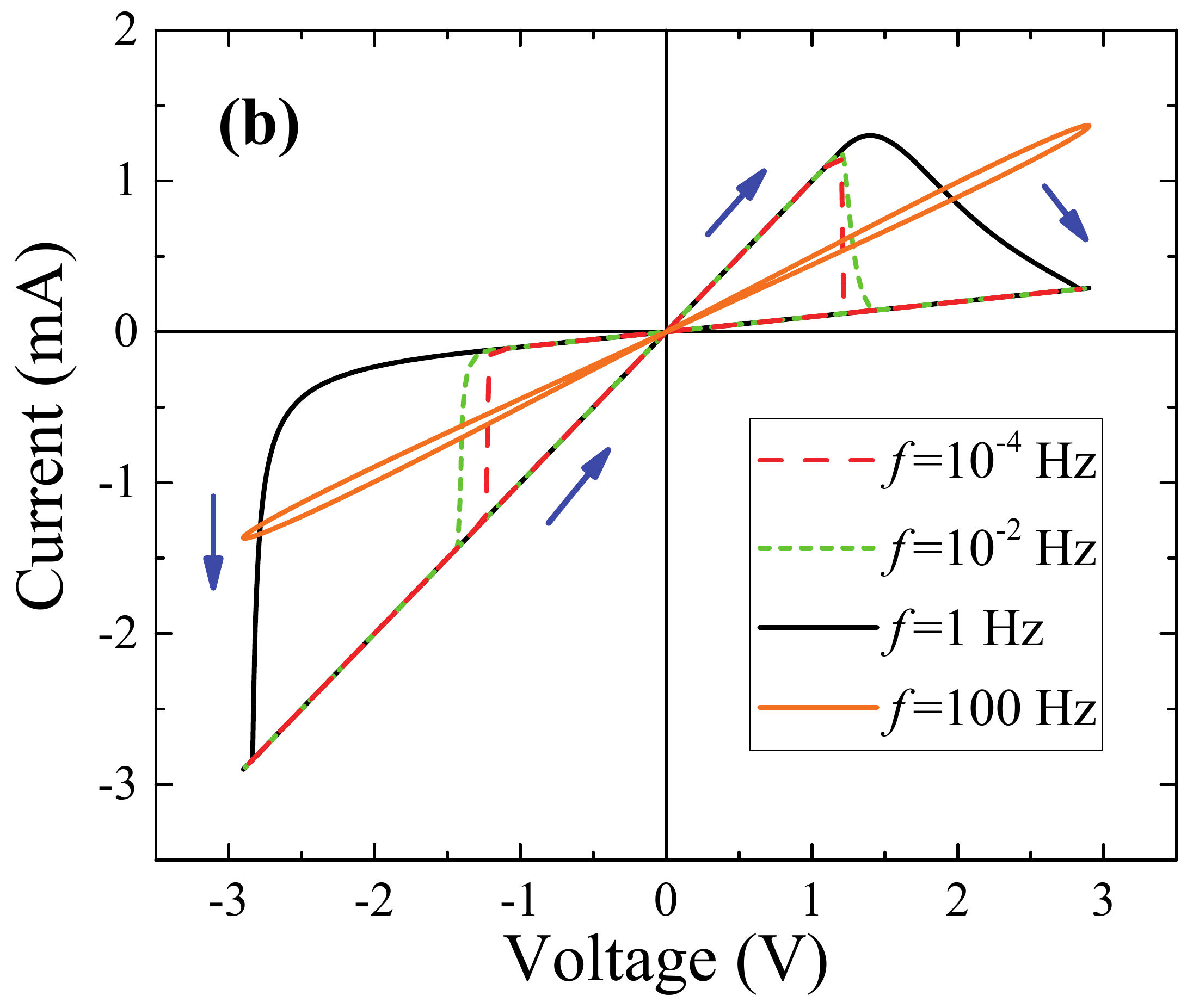}
\caption{(Color online) (a) A particular realization of a clock signal
generator. This circuit employs an OR gate with Schmitt-trigger inputs (to
the left), and open-drain identity (upper) and NOT (lower) gates. Here $V_{%
\mathrm{p}}=5$~V is the power supply voltage. (b) $I-V$ curves of the
memristive system M used in this work. The memristive system parameters are
given in the text.}
\label{fig2}
\end{figure*}

Here we demonstrate a new miniaturized clock signal generator based on a
memristive (memory resistive) device \cite{chua76a} operating in a
Sisyphus-like cycle. In modern electronics, the clock signal is most
frequently produced by quartz generators and sometimes by $RC$-based
circuits or other approaches. The quartz generators offer a high precision
at the cost of their size. Less precise $RC$-circuits are more compact.
Nanoscale memristive devices \cite{pershin11a} [combining their very small
size with switching frequencies in a convenient range (e.g., hundreds of
MHz)] are the core components of clock generators explored here.

In the literature, there are several known methods of using memristors in oscillating circuits.
In particular, significant attention has been focused on nonlinear oscillators constructed
from Chua's oscillators, by replacing Chua's diodes with memristors \cite{Itoh08a,Corinto11a}.
The authors of  Ref.~\citeonline{pershin09d} proposed a programmable frequency-relaxation oscillator.
In such a circuit, a memristor-based digital potentiometer is used to set switching thresholds of a Schmitt
trigger. Moreover, memristors can be employed to replace capacitors in relaxation oscillators resulting in
compact reactance-less oscillators \cite{Zidan11a,Mosad13a,Yu14a,Zidan14a,Kyriakides15a}. Additionally,
it was shown experimentally that some polymeric memory devices exhibit slow current oscillations when
subjected to a constant voltage \cite{Erokhin07b}.  Our work presents an advanced reactance-less oscillator
design having a unique combination of only digital logic gates, a single-supply voltage
and a realistic threshold-type memristive device.


In the proposed clock signal generator, the frequency is defined by the
switching characteristics of a memristive device. Our circuit employs
bi-polar memristive devices \cite{pershin11a} operating in such a way that
their memristances (memory resistances) increase/decrease at
positive/negative voltages applied to the device, respectively. The
effective circuits depicted in Fig.~\ref{fig1}(d) show the two stages of its
operation. In stage 1 (left circuit), the positive voltage applied to the
memristor M causes an increase in its memristance. In the second stage, the
device polarity is changed and the memristance decreases. In this way, the
memristance oscillates between two values [see Fig.~\ref{fig1}(e)],
similarly to the oscillations of the boulder height above the ground level
in the Sisyphus case [Fig.~\ref{fig1}(c)]. Importantly, the circuit does not
require any large-size components (such as a quartz resonator) typically
used in conventional oscillator circuits.


\section*{Memristive clock signal generator}

Figure~\ref{fig2}(a) presents the specific memristive clock generator
circuit introduced in this work. Its components include the memristive
system M, an OR gate with Schmitt-trigger inputs (to the left), open-drain
identity (upper) and NOT (lower) gates, three resistors and a capacitor.
Note that the resistors $R_{1}$ and $R_{2}$ here play the same role as $R_{1}
$ and $R_{2}$ in the effective circuits shown in Fig.~\ref{fig1}(e). As will
be readily observed, the output of the OR gate defines the operation stages:
logical $0$ corresponds to the increasing memristance stage $1$, while
logical $1$ corresponds to the decreasing memristance stage $2$. The
hysteretic input levels of the Schmitt-trigger inputs (denoted by $V_{+}$
for the logic $1$, and $V_{-}$ for the logic $0$, $V_{-}<V_{+}$) are
employed as voltage thresholds triggering the stage changes. Figure~\ref%
{fig2}(b) shows the calculated pinched hysteresis loops of the memristive
system M described by Eqs.~(\ref{eq:model1}-\ref{eq:model2}). In this
calculation, $R_{\mathrm{M}}(t=0)=2$\thinspace k$\Omega $, $V_{\mathrm{M}%
}(t)=V_{0}\sin (2\pi ft)$, $V_{0}=2.9$\thinspace V, and all other parameters
of M are the same as in the memristor emulator specified below Eqs. (\ref%
{eq:model1})-(\ref{eq:model2}).

One can notice from Fig.~\ref{fig2}(a) that indeed, the increasing
memristance stage $1$ switches to the decreasing memristance stage $2$ as
soon as the voltage level $V_{\mathrm{M}}$ reaches the $V_{+}$ threshold. At
this instance in time, the output of the OR gate changes to 1, grounding the
bottom terminal of M and setting the identity gate into the high impedance
state. Assuming that $V_{\mathrm{M},2}>V_{-}$ right after the switching
(below we discuss all the related requirements), the circuit will remain in
the decreasing memristance stage $2$ while $V_{\mathrm{M},2}>V_{-}$. The
transition from stage $2$ to stage $1$ occurs in a similar way (as soon as $%
V_{\mathrm{M},2}<V_{-}$, and thus both inputs of OR become logical zeros).
In Fig.~\ref{fig2}(a), the $R_{3}C$ circuit introduces a short time delay to
ensure proper switching between the phases. The generator output frequency
is not influenced by this delay.

Next, we discuss the experimental implementation of the clock signal
generator. In our experiments, the memristive system M is realized with a digital memristor emulator \cite%
{pershin09d,pershin2012teaching}.  Its
main parts include a microcontroller, analog-to-digital converter and
digital potentiometer. The operation of the digital memristor emulator is straightforward:
using the analog-to-digital converter, the microcontroller cyclically measures the
voltage applied to the digital potentiometer, calculates an
updated value of the memristance (using pre-programmed
equations of voltage-controlled or current-controlled memristive
system \cite{chua76a}), and writes the updated value of the memristance into the digital potentiometer.
Figure~5(b) of Ref.~\citeonline{pershin2012teaching} presents a photograph of the specific digital memristor emulator
realization employed in the present paper.
More details regarding the emulator design can be found in Refs.~\citeonline{pershin09d,pershin2012teaching}.

The memristor emulator was pre-programmed with a model of
voltage-controlled memristive systems with threshold \cite%
{pershin13a,pershin09b}
\begin{eqnarray}
I &=&R_{\mathrm{M}}^{-1}(x)V_{\mathrm{M}},  \label{eq:model1} \\
\frac{\mathrm{d}x}{\mathrm{d}t} &=&\begin{cases} \pm
\textnormal{sign}(V_{\mathrm{M}})\beta
(|V_{\mathrm{M}}|-V_{\mathrm{t}})\;\textnormal{if}\;\;|V_{\mathrm{M}}|>V_{%
\mathrm{t}} \\ 0\;\;\;\;\;\;\;\;\;\;\;\;\textnormal{otherwise}\end{cases},
\label{eq:model2}
\end{eqnarray}%
where $I$ and $V_{\mathrm{M}}$ are the current through and the voltage
across the memristive system, respectively, and $x$ is the internal state
variable playing the role of the memristance. Here $R_{\mathrm{M}}(x)\equiv x
$, $\beta $ is a positive switching constant characterizing the intrinsic
rate of memristance change when $|V_{\mathrm{M}}|>V_{\mathrm{t}}$, $V_{%
\mathrm{t}}$ is the threshold voltage, and the $+$ or $-$ sign is selected
according to the device connection polarity. Additionally, it is assumed
that the memristance is limited to the interval [$R_{\mathrm{on}}$, $R_{%
\mathrm{off}}$] (note that $R_{\mathrm{on}}<R_{\mathrm{off}}$). The specific
model parameters used in our emulator are $\beta =62$~k$\Omega /$V$\cdot $s,
$V_{\mathrm{t}}=1.2$~V, $R_{\mathrm{on}}=1$~k$\Omega $, and $R_{\mathrm{off}%
}=10$~k$\Omega $, and $R_{\mathrm{M}}(t=0)=(R_{\mathrm{on}}+R_{\mathrm{off}})/2$. We built the circuit shown in Fig.~\ref{fig2}(b) using a
TI SN74HC7032 positive-OR gate with Schmitt-trigger inputs, CD74AC05
inverters with open-drain outputs, $R_{1}=4.7$~k$\Omega $, $R_{2}=2.7$~k$%
\Omega $, $R_{3}=10$~k$\Omega $, and $C=1.35$~nF.

\begin{figure}[t]
\centering
\includegraphics[width=.43\columnwidth]{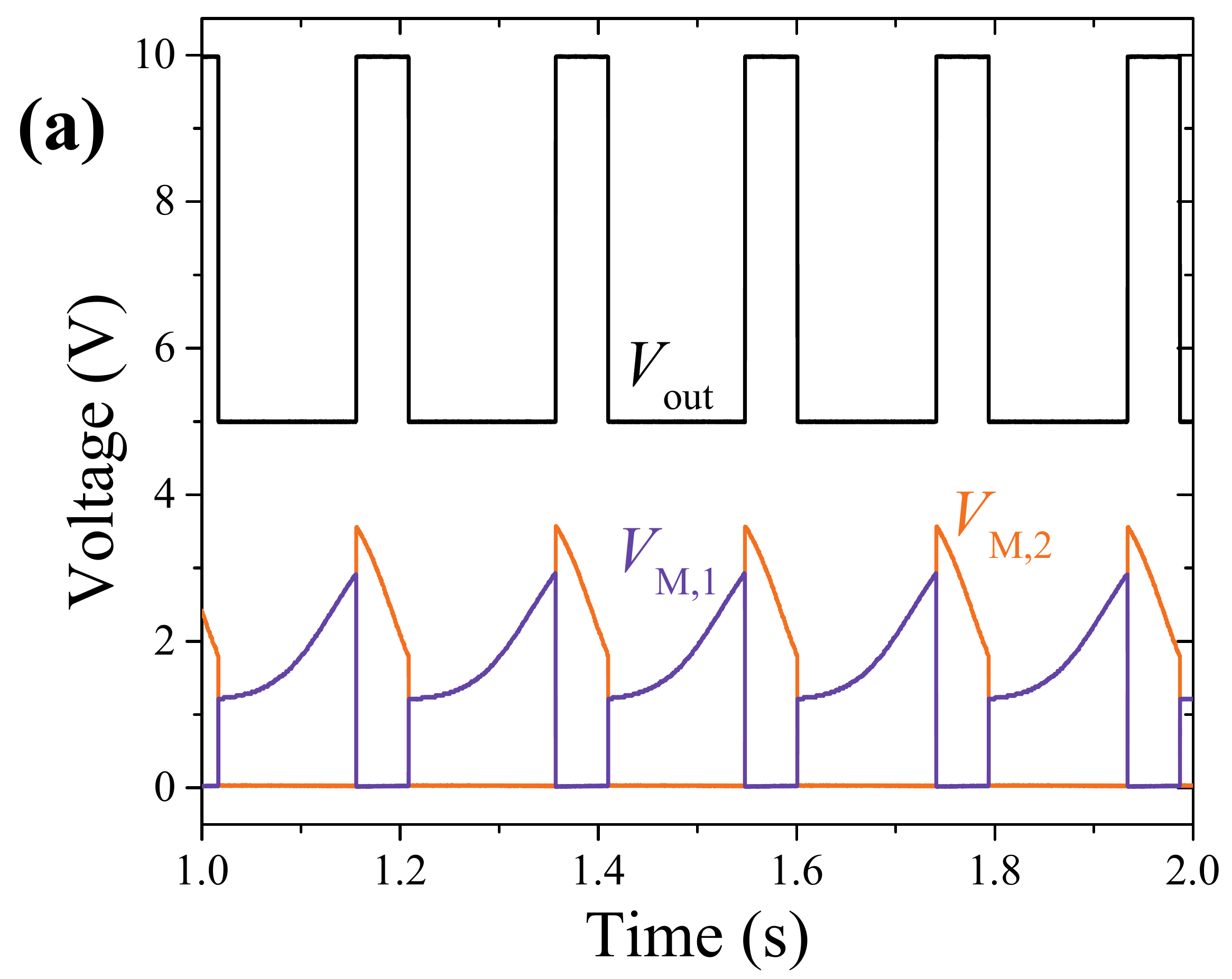} \quad \quad \quad \quad %
\includegraphics[width=.42\columnwidth]{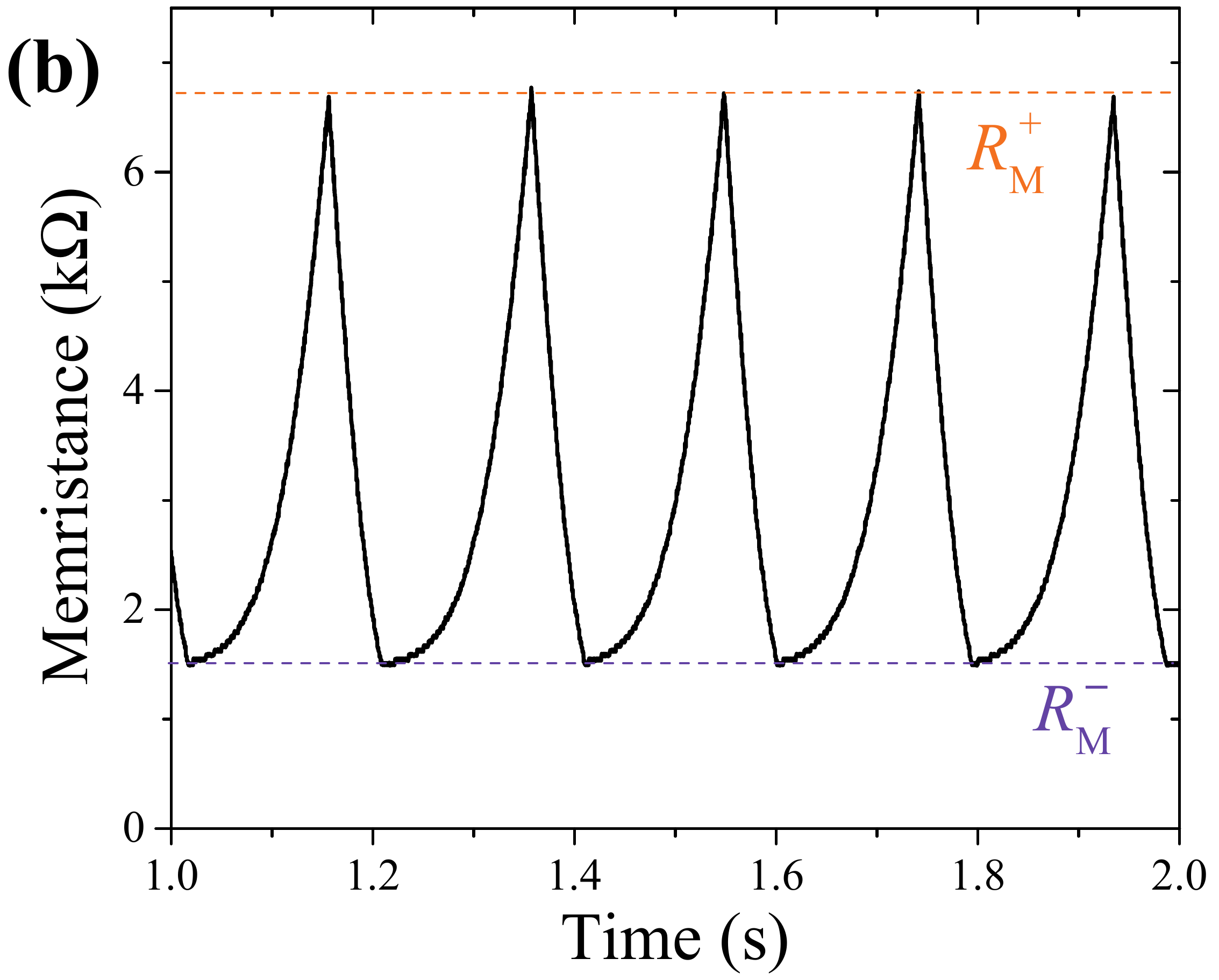}
\caption{(Color online) (a) Experimentally measured voltages at the output
of the OR gate ($V_{\mathrm{out}}$) and top and bottom electrodes of the
memristive system ($V_{\mathrm{M},2(1)}$) in the circuit shown in Fig.~%
\protect\ref{fig2}(a). For clarity, the $V_{\mathrm{out}}$ is displaced by $%
5~$V. (b) Memristance $R_{\mathrm{M}}$ extracted from the data presented in
(a) by using Eq.~(\protect\ref{eq:3}).}
\label{fig3}
\end{figure}

Our circuit is fully reproducible within the specifications of the circuit components. Assuming that the future memristor devices will be available with well characterized characteristics (similarly, for example, to usual resistors or capacitors), we expect that the suggested circuit based on real memristors will be reproducible too. In addition, our measurements and simulations did not display any significant dispersion of the memristive properties as well of the output signal produced by the circuit. Basically, the memristor emulator operates deterministically and thus the circuit also operates deterministically. As deterministic models are frequently used to describe the response of real memristive devices and, in many cases, show a very good agreement with experiments, we believe that the selected approach (based on the emulator) illustrates our idea in a very realistic manner.

Figure \ref{fig3}(a) presents results of our measurements. While the $V_{%
\mathrm{M},1}$ and $V_{\mathrm{M},2}$ curves clearly demonstrate two stages
of the circuit operation, the output $V_{\mathrm{out}}$ [displayed in Fig.~%
\ref{fig2}(a)] shows a stable clock signal with a period of about
0.194\thinspace s and a duty cycle of about 27\%. With the knowledge of $%
R_{1}$ and $R_{2}$, the $V_{\mathrm{M},1}$ and $V_{\mathrm{M},2}$ curves
were used to extract the time dependence of the memristance $R_{\mathrm{M}}$
depicted in Fig.~\ref{fig3}(b). Its variations are similar to the desired
variations of $R_{\mathrm{M}}$ sketched in Fig.~\ref{fig1}(e).

\section*{Circuit analysis}

In Fig.~\ref{fig3}(b), the memristance $R_{\mathrm{M}}$ of M changes
periodically from $R_{\mathrm{M}}^{-}$ to $R_{\mathrm{M}}^{+}$, and back
(see also Fig.~\ref{fig1}(e)). Analytically, one can find the corresponding
durations of the time intervals, $\tau _{1}$ and $\tau _{2}$. In what
follows, we focus on the effective circuit models depicted in Fig.~\ref{fig1}%
(d) to find $\tau _{1}$ and $\tau _{2}$ analytically.

\begin{figure}[th]
\centering
\includegraphics[width=.65\columnwidth]{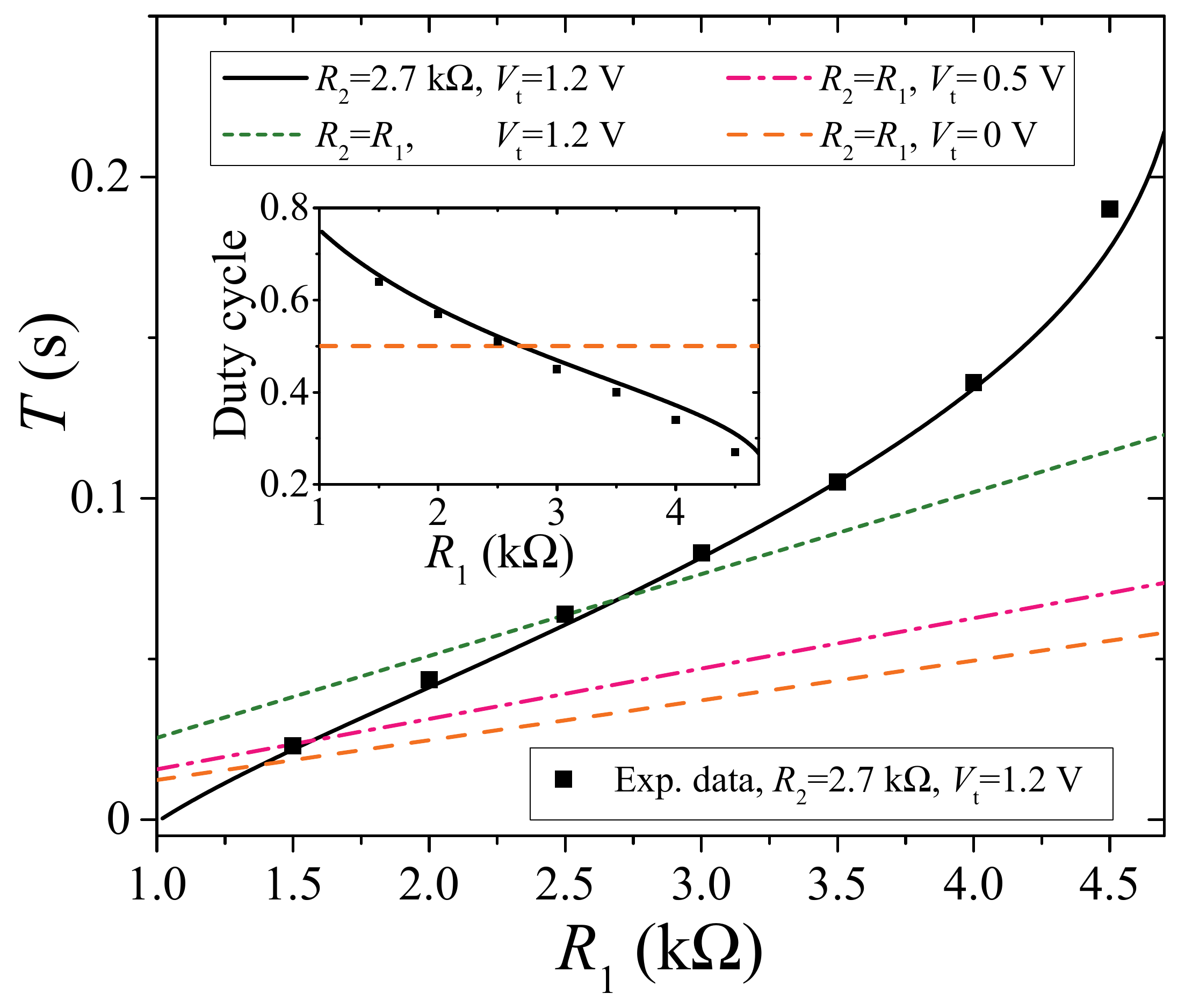}
\caption{(Color online) Period $T=\protect\tau _{1}+\protect\tau _{2}$ as
function of the resistance $R_{1}$. Inset: the duty cycle $\protect\tau %
_{2}/T$ as a function of $R_{1}$. This plot was obtained using the following
set of parameters: $V_{\mathrm{p}}=5$~V, $V_{+}=3$~V, $V_{-}=1.8$~V, and $%
\protect\beta =62$ k$\Omega /$V$\cdot $s. Note that $R_{2}$ and $V_{\mathrm{t%
}}$ are indicated on the plot. }
\label{fig5}
\end{figure}

For the sake of convenience, let us consider the system dynamics from $t=0$
in both stages and perform calculations based on the memristive device model
employed in our emulator [Eqs.~(\ref{eq:model1}, \ref{eq:model2})]. Then,
using
\begin{equation}
V_{\mathrm{M},i}(t)=V_{\mathrm{p}}\frac{R_{\mathrm{M},i}(t)}{R_{i}+R_{%
\mathrm{M},i}(t)},  \label{eq:3}
\end{equation}%
where $i=1,2$ denotes the two time intervals, we integrate Eq.~(\ref%
{eq:model2}) with appropriate initial and final conditions for both
intervals, and accounting for the memristor polarity. As a result, one can
find the following expression for the duration of the two stages
\begin{equation}
\tau _{i}=\frac{1}{\beta }\frac{R_{\mathrm{M}}^{+}-R_{\mathrm{M}}^{-}}{V_{%
\mathrm{p}}-V_{\mathrm{t}}}+\frac{1}{\beta }\frac{R_{i}V_{\mathrm{p}}}{%
\left( V_{\mathrm{p}}-V_{\mathrm{t}}\right) ^{2}}\ln \left\{ \frac{\left( V_{%
\mathrm{p}}-V_{\mathrm{t}}\right) R_{\mathrm{M}}^{+}-V_{\mathrm{t}}R_{i}}{%
\left( V_{\mathrm{p}}-V_{\mathrm{t}}\right) R_{\mathrm{M}}^{-}-V_{\mathrm{t}%
}R_{i}}\right\} .  \label{eq:tau}
\end{equation}%
The oscillation period is given by
\begin{equation}
T=\tau _{1}+\tau _{2}.  \label{eq:period}
\end{equation}

The boundary values of the memristance, $R_{\mathrm{M}}^{-}$ and $R_{\mathrm{%
M}}^{+}$, can be expressed through the Schmitt-trigger input thresholds as
\begin{equation}
R_{\mathrm{M}}^{+}=\frac{R_{1}V_{+}}{V_{\mathrm{p}}-V_{+}},\quad \quad R_{%
\mathrm{M}}^{-}=\frac{R_{2}V_{-}}{V_{\mathrm{p}}-V_{-}}.  \label{Rpm}
\end{equation}

The expression for the period $T$ can be written in a simple form assuming
that $R_{1}=R_{2}$ and $V_{\mathrm{p}}\gg V_{\mathrm{t}}$ (and also $%
R_{1}\lesssim R_{\mathrm{M}}^{-}$ so that $V_{\mathrm{p}}R_{\mathrm{M}%
}^{-}\gg V_{\mathrm{t}}R_{1}$). Then we obtain%
\begin{equation}
T=\frac{2}{\beta }\frac{R_{1}}{V_{\mathrm{p}}}\left\{ \frac{R_{\mathrm{M}%
}^{+}-R_{\mathrm{M}}^{-}}{R_{1}}+\ln \frac{R_{\mathrm{M}}^{+}}{R_{\mathrm{M}%
}^{-}}\right\} .  \label{T_}
\end{equation}%
This together with Eq.~(\ref{Rpm}) gives%
\begin{equation}
T=\frac{2}{\beta }\frac{R_{1}}{V_{\mathrm{p}}}\left\{ \frac{V_{\mathrm{p}%
}\left( V_{+}-V_{-}\right) }{\left( V_{\mathrm{p}}-V_{+}\right) \left( V_{%
\mathrm{p}}-V_{-}\right) }+\ln \frac{V_{+}\left( V_{\mathrm{p}}-V_{-}\right)
}{V_{-}\left( V_{\mathrm{p}}-V_{+}\right) }\right\} .  \label{T__}
\end{equation}

The above formulas are plotted in Fig.~\ref{fig5}. There, it is assumed that
all the circuit parameters (except~$R_{1}$ and $R_{2}$) are fixed. This
figure shows that both the period $T$ and the duty cycle $\tau _{2}/T$ can
be tuned in a certain range by varying $R_{1}$ and/or $R_{2}$. In Fig.~\ref%
{fig5}, the solid curve was plotted for the same value of $R_{2}$ as used in
the experiment, while the other curves demonstrate the changes assuming
equal resistances, $R_{2}=R_{1}$, for several values of the memristor
threshold voltage $V_{\mathrm{t}}$. In particular, Fig.~\ref{fig5}
demonstrates that the formula~(\ref{T__}), which is valid at $V_{\mathrm{t}%
}=0$ and $R_{2}=R_{1}$, provides a good estimation for the period $T$ for
small values of the threshold voltage $V_{\mathrm{t}}$ when $R_{2}=R_{1}$.
Moreover, Fig. 4 also shows several experimentally measured periods and corresponding
duty cycles that exhibit a very good agreement with our analytical results.

Finally, we consider the limitations imposed on the clock signal generator
components required for its proper operation. The numerical estimations
provided below employ parameter values that are close to those
of our experimental implementation of the signal generator. For the convenience of readers, here
we list the values of these parameters: $V_{\mathrm{p}}=5$~V, $V_{+}=3$~V,
$V_{-}=1.8$~V, $V_{\mathrm{t}}=1.2$~V, $R_{\mathrm{on}}=1$~k$\Omega ,$ and $%
R_{\mathrm{off}}=10$~k$\Omega$.

\begin{enumerate}
\item \label{cr1} In the stage $1$, the minimum (maximum) values of $V_{%
\mathrm{M}}$ must be below (above) $V_{+}$; namely, $V_{\mathrm{M}}(R_{%
\mathrm{on}})<V_{+}$ and $V_{\mathrm{M}}(R_{\mathrm{off}})>V_{+}$.
Consequently,
\begin{equation}
R_{\mathrm{on}}\frac{\left( V_{\mathrm{p}}-V_{+}\right) }{V_{+}}<R_{1}<R_{%
\mathrm{off}}\frac{\left( V_{\mathrm{p}}-V_{+}\right) }{V_{+}}.
\label{eq:cond1}
\end{equation}%
Numerically, $0.667$~k$\Omega $ $<R_{1}<6.67$~k$\Omega $.

\item \label{cr2} In the stage $2$, the minimum (maximum) values of $V_{%
\mathrm{M}}$ must be below (above) $V_{-}$, namely, $V_{\mathrm{M}}(R_{%
\mathrm{on}})<V_{-}$ and $V_{\mathrm{M}}(R_{\mathrm{off}})>V_{-}$.
Consequently,
\begin{equation}
R_{\mathrm{on}}\frac{\left( V_{\mathrm{p}}-V_{-}\right) }{V_{-}}<R_{2}<R_{%
\mathrm{off}}\frac{\left( V_{\mathrm{p}}-V_{-}\right) }{V_{-}}.
\label{eq:cond2}
\end{equation}%
Numerically, $1.778$~k$\Omega $ $<R_{2}<17.78$~k$\Omega $.

\item \label{cr3} Moreover, right after the switching from stage $1$ to
stage $2$, the voltage across M should stay above $V_{-}$, namely, $V_{-}<V_{%
\mathrm{p}}R_{\mathrm{M}}^{+}/\left( R_{\mathrm{M}}^{+}+R_{2}\right) $,
where $R_{\mathrm{M}}^{+}$ is given by Eq.~(\ref{Rpm}). It follows that
\begin{equation}
\frac{R_{2}}{R_{1}}<\frac{V_{+}\left( V_{\mathrm{p}}-V_{-}\right) }{%
V_{-}\left( V_{\mathrm{p}}-V_{+}\right) }  \label{eq:cond3}
\end{equation}%
and, numerically, $R_{2}/R_{1}<2.67$. We note that the requirement that in
the transition from stage $2$ to stage $1$ the voltage across M stays below $%
V_{+}$, is also given by Eq.~(\ref{eq:cond3}). For the same transition, one
can also require that $V_{\mathrm{M}}(R_{\mathrm{M}}^{-})>V_{\mathrm{t}}$ in
stage $1$, however, this requirement is weaker than the criterion \ref{cr4}.

\item \label{cr4} In order to start oscillations from \emph{any} initial
condition, the voltage across M (in the worst-case limit $R_{\mathrm{M}}=R_{%
\mathrm{on}}$) should exceed $V_{\mathrm{t}}$. This results in
\begin{equation}
R_{1}<R_{\mathrm{on}}\left( \frac{V_{\mathrm{p}}}{V_{\mathrm{t}}}-1\right) .
\label{eq:cond4}
\end{equation}%
Using the above-mentioned parameters, we obtain $R_{1}<3.167$~k$\Omega $.

\item \label{cr5} Lastly, we mention the obvious requirement on $V_-$,
namely, $V_{\mathrm{t}}<V_{-}$.
\end{enumerate}

One can easily notice that in our experimental implementation of the circuit
shown in Fig.~\ref{fig2}(b), the selected values of $R_{1}=4.7$~k$\Omega $
and $R_{2}=2.7$~k$\Omega $ satisfy the criteria in Eqs. (\ref{eq:cond1}-\ref%
{eq:cond3}). Regarding the criterion (\ref{eq:cond4}), we were able to use a
large value of $R_{1}$, because the initial value of the emulator
memristance was selected above $R_{\mathrm{on}}$.

\section*{Conclusion}

We have proposed and analyzed the design of a memristive
clock signal generator. Assuming a realistic threshold model of a
memristive device, we experimentally demonstrated the operation of a
memristive frequency generator. While our demonstration is based on a slow
memristor emulator, the real memristive devices can result in frequencies in
the industrially important MHz-GHz range. Moreover, the specific proposed
circuit for frequency generation offers frequency and duty cycle tunability.
These theoretical considerations together with our experimental emulation
shows the potential of such circuits for very compact frequency generators.

\section*{Acknowledgements}

This work has been supported by the NSF grant No. ECCS-1202383, USC Smart
State Center for Experimental Nanoscale Physics, the RIKEN iTHES Project, MURI
Center for Dynamic Magneto-Optics, and a Grant-in-Aid for Scientific
Research (A), and Russian Scientific Foundation grant No. 15-13-20021.


\end{document}